 \documentclass[final,3p,times]{elsarticle}

\setcitestyle{square}
\usepackage{enumitem}
\usepackage[utf8]{inputenc}
\usepackage[english]{babel}
\usepackage[usenames, dvipsnames]{color}

 \usepackage[numbers]{natbib}
\usepackage{float}
\usepackage{graphicx,dblfloatfix}
\usepackage{subfig}
\usepackage[numbers]{natbib}

\usepackage[cmex10]{amsmath}
\usepackage{array}

\DeclareMathOperator*{\amin}{min}

\let\oldtabular\tabular
\renewcommand{\tabular}{\footnotesize\oldtabular}

\journal{}

\begin{document}

\begin{frontmatter}


 \title{Optimal Layered Representation for Adaptive Interactive Multiview Video Streaming}
 
 \author[epfl,ist]{Ana~De~Abreu \corref{cor1}}
 \ead{ana.deabreu@epfl.ch}
 \author[epfl]{Laura~Toni}
 \ead{laura.toni@epfl.ch}
 \author[essex]{Nikolaos~Thomos}
 \ead{nthomos@essex.ac.uk}
 \author[inria]{Thomas~Maugey}
 \ead{thomas.maugey@inria.fr}
 \author[ist]{Fernando~Pereira }
 \ead{fp@lx.it.pt}
 \author[epfl]{Pascal~Frossard}
 \ead{pascal.frossard@epfl.ch}
 
 \cortext[cor1]{Corresponding author}


 \address[epfl]{Signal Processing Laboratory (LTS4), Ecole Polytechnique F\'{e}d\'{e}rale de Lausanne (EPFL), CH-1015 Lausanne, Switzerland.}
   \address[ist]{Instituto Superior T\'{e}cnico, Universidade de Lisboa - Instituto de Telecomunica\c{c}\~{o}es (IST/UL-IT), 1049-001, Lisbon, Portugal.}
 \address[essex]{University of Essex, Colchester, United Kingdom.}
  \address[inria]{Inria Rennes Bretagne Atlantique,  Campus de Beaulieu,  35042 Rennes Cedex, France.}

\begin{abstract}
We consider an interactive multiview video streaming (IMVS) system where clients select their preferred viewpoint in a given navigation window. To provide high quality IMVS, many high quality views should be transmitted to the clients. However, this is not always possible due to the limited and heterogeneous capabilities of the clients. In this paper, we propose a novel adaptive IMVS solution based on a \emph{layered multiview representation} where camera views are organized into layered subsets to match the different clients constraints. We formulate an optimization problem for the joint selection of the views subsets and their encoding rates. Then, we propose an optimal and a reduced computational complexity greedy algorithms, both based on dynamic-programming. Simulation results show the good performance of our novel algorithms compared to a baseline algorithm, proving that an effective IMVS adaptive solution should consider the scene content and the client capabilities and their preferences in navigation.

\end{abstract}

\begin{keyword}
Interactive multiview video, layered representation, navigation window, view synthesis.

\end{keyword}

\end{frontmatter}

\section{Introduction}\label{sec:introduction}

In emerging multiview video applications an array of cameras captures the same 3D scene from different viewpoints in order to provide the clients with the capability of choosing among different views of the scene. Intermediate virtual views, not available from the set of captured views, can also be  rendered at the decoder by depth-image-based rendering (DIBR) techniques \cite{Schmeing2011}, if texture information and depth  information of neighboring views are available. As a result, \emph{interactive multiview video} clients have the freedom of selecting a viewpoint from a set of captured and virtual views that define a navigation window. The quality of the rendered views in the navigation window depends on the quality of the captured views and on their relative distance, as the distortion of a virtual view tend to increase with the distance to the views used as references in the view synthesis process \cite{Ana2014}. This means that in the ideal case, all the captured views, encoded at the highest possible rate, would be transmitted to all the clients. However, in practice,  resource constraints prevent the transmission of all the views. In particular, clients may have different access link bandwidth capabilities, and some of them may not be able to receive all the captured views. In this context, it becomes important to devise adaptive transmission strategies for interactive multiview video streaming (IMVS) systems that adapt to the capabilities of the clients.

\begin{figure}[h]
\centering
\includegraphics[width=3.1in]{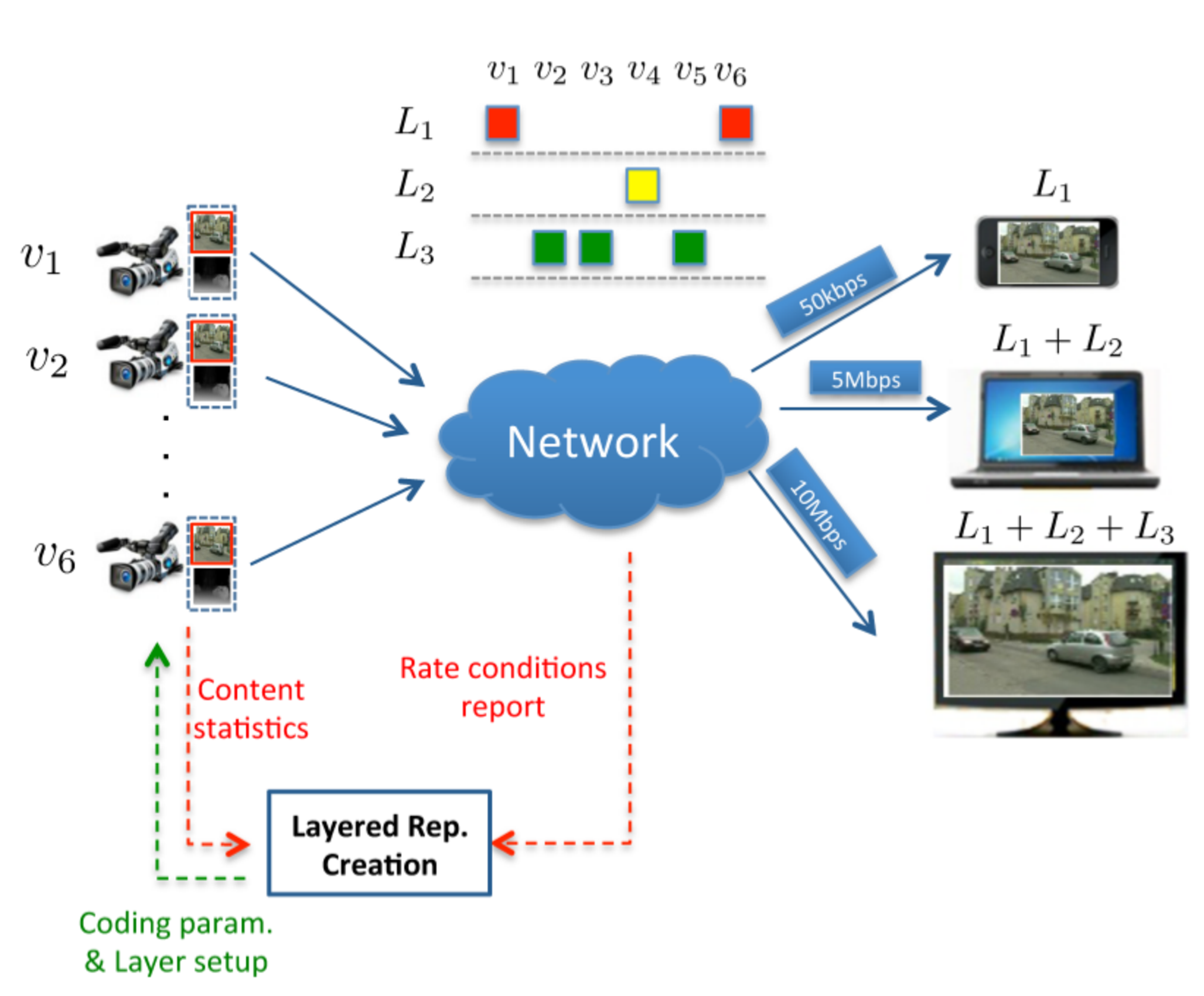}
\caption{Illustration of an IMVS system with 6 camera views and 3 heterogeneous clients. The optimization is done by the \emph{layered representation creation} module considering three layers. } \label{fig:Nmodel}
\end{figure}

 In this work, we consider the problem of jointly determining which views to transmit and at what encoding rate, such that the expected rendering quality in the navigation window is maximized under relevant resource constraints. In particular, we consider the scenario illustrated in Fig. \ref{fig:Nmodel}, where a set of views are captured from an array of time-synchronized cameras. For each captured view, both a texture and a depth map are available, so that intermediate virtual viewpoints can eventually be synthesized. The set of captured and virtual views defines the navigation window available for client viewpoint request. Clients are clustered according to their bandwidth capabilities; for instance, in Fig. \ref{fig:Nmodel} only one client per cluster is illustrated for three groups with 50kbps, 5Mbps and 10Mbps bandwidth constraints. Then, the set of captured views are organized in layers or subsets of views to be transmitted to the different groups of clients in order to maximize the overall navigation quality. With a layered organization of the captured views in the navigation window, we aim at offering a progressive increase of the rendering quality, as the quality of the navigation improves with the number of layers (subset of views) that clients are able to receive. In the example of Fig. \ref{fig:Nmodel},  three layers or subsets of views are formed as: ${L}_1=\{v_1, \; v_6\}$, ${L}_2=\{v_4\}$ and ${L}_3=\{v_2, \; v_3, \; v_5\}$. Depending on the clients' bandwidth capabilities, they receive the views in ${L}_1$, or in ${L}_1$ and ${L}_2$, or in ${L}_1$, ${L}_2$ and ${L}_3$. In particular, the client with the lowest bandwidth capability (the client with a mobile phone) is able to receive only the subset of views $\{v_1, \; v_6\}$, and needs to synthesize the rest of the views. On the other hand, the client with the highest bandwidth capability (the client with a TV), is able to receive all the views, and therefore reaches the highest navigation quality.

We formulate an optimization problem to jointly determine the optimal arrangement of views in layers along with the coding rate of the views, such that the expected rendering quality is maximized in the navigation window, while the rate of each layer is constrained by network and clients capabilities. We show that this combinatorial optimization problem is NP-hard, meaning that it is computationally difficult and there are not known algorithm that optimally solves the problem in polynomial time. We then propose a globally optimal solution based on dynamic-programing (DP) algorithm. As the computational complexity of this algorithm grows with the number of layers, a greedy and lower complexity algorithm is proposed, where the optimal subset of views and their coding rates are computed successively for each layer by a DP-based approach. The results show that our greedy algorithm achieves a close-to-optimal performance in terms of total expected distortion, and outperforms a distance-based view and rate selection strategy used as a baseline algorithm for layer construction.

This paper is organized as follows. First, Section \ref{sec:related_work} discusses the related work. Then, the main characteristics of the layered interactive multiview video representation are outlined in Section \ref{sec:framework} where also our optimization problem is formulated. Section \ref{sec:GLA} describes the optimal and greedy views selection and rates allocation algorithms for our layered multiview representation. Section \ref{sec:results}  presents the experimental results that show the benefits of the proposed solution and the conclusions are outlined in Section \ref{sec:conclusions}.


\section{Related work}\label{sec:related_work}

In this section, we review the work related to the design of IMVS systems by focusing on the problem of data representation and transmission in constrained resources environments. 

In general, the limited bandwidth problem in IMVS  has been approached by proposing some coding/prediction structure optimization mechanisms for the compression of multiview video data. In \cite{Gene2008}, \cite{Gene2011} and \cite{Gene2012}, the authors have studied the prediction structures based on redundant P- and DSC-frames (distributed source coding) that facilitate a continuous view-switching by trading off the transmission rate and the storage capacity. To save transmission bandwidth, different interview prediction structures are proposed in \cite{Liu2010} to code in different ways each multiview video dataset, in order to satisfy different rate-distortion requirements. In \cite{Ana2013}, \cite{Ana2014_journal} and  \cite{Ana2013_2}, a prediction structure selection mechanism has been proposed for minimal distortion view switching while trading off transmission rate and storage cost in the IMVS system. 

A different coding solution to the limited bandwidth problem has been proposed in \cite{Takuya2014}, where a \emph{user dependent multiview video streaming for Multi-users} (UMSM) system is presented to reduce the transmission rate due to redundant transmission of overlapping frames in multi-user systems. In UMSM, the overlapping frames (potentially requested by two or more users) are encoded together and transmitted by multicast, while the non-overlapping frames are transmitted to each user by unicast. Differently, the authors in \cite{Gene2011_2} and  \cite{Gene2010} tackle the problem of scarce transmission bandwidth by determining the best set of camera views for encoding and by efficiently distributing the available bits among texture and depth maps of the selected views, such that the visual distortion of reconstructed views is minimized given some rate constraints.  

Although these works propose solutions to the constrained bandwidth problem in IMVS, they do not consider the bandwidth heterogeneity of the clients, and rather describe solutions that do not adapt to the different capabilities of the clients.

The adaptive content concept in multiview video has been mostly used in the coding context, where the problem of heterogeneous clients has been tackled via scalable multiview video coding. For instance, some extensions of the H.264/SVC standard \cite{SVC} for traditional 2D video have been proposed in the literature for multiview video \cite{Drose2006} \cite{Ozbek2006}. In  \cite{Gene2011_3}, \cite{Jacob2013} and\cite{Jacob2015}, the authors propose a joint view and rate adaptation solution for heterogeneous clients. Their solution is based on a wavelet multiview image codec that produces a scalable bitstream from which different subsets can be extracted and decoded at various bitrates in order to satisfy different clients bandwidth capabilities.

In addition, multiview video permits the introduction of a new type of adaptive content compared to classical video. For instance, instead of transmitting the complete set of views of the multiview video dataset, some views can be omitted from the compressed bitstream and eventually reconstructed at the receiver side using DIBR method. The more the transmitted views are, the higher the reconstruction quality is but the larger the bitrate is too. This new type of adaptive content that trades off navigation quality and transmission bandwidth, has been studied in  \cite{Ana2014} and \cite{Laura2013}. In these works, the set of captured views are organized in subsets, that we call layers, and they are transmitted to clustered heterogeneous clients according to their bandwidth capabilities.

The work in \cite{Laura2013}, however does not optimize the set of views per layer, but rather distribute them based on a uniform distance criteria. The work in \cite{Ana2014} optimizes the selection of views in particular settings, but does not adapt the coding rate of each camera view. In this paper, we build on these previous works and extend the state-of-the-art by proposing solutions for jointly optimizing the selection of the views and their coding rate in a layered data representation for IMVS systems, such that the expected rendering quality is maximized in the navigation window, while the rate of each layer is constrained by network and clients capabilities. To solve this problem, we propose an optimal algorithm and a greedy algorithm with a reduced complexity, both based on dynamic-programming. These algorithms  adapt to the client capabilities and their preferences in navigation, to the camera positions and to the content of the 3D scene, in order to have an effective addaptive solution for IVMS systems.


\section{Framework and Problem Formulation} \label{sec:framework}

We consider the problem of building a layered multiview video representation in an IMVS system, where the clients are heterogeneous in terms of bandwidth capabilities. In this section, we first describe the most relevant characteristics of the IMVS system. Then, we formally formulate our optimization problem.


\subsection{Network and IMVS model} \label{subsec:Nmodel}

In this work, we define $\mathcal{V}=\{v_1, v_2, \ldots, v_V \}$  as the ordered set of captured views from an array of  cameras. Each camera compresses the recorded view before transmitting it over the network. We assume that there is no communication between the cameras, so each camera encodes its images independently of the other cameras, which is common in numerous novel applications ranging from surveillance to remote sensing. For each captured view $v_j \in \mathcal{V}$, both a texture image (image containing color information) and a depth map (image containing distance information to objects in the 3D scene) are captured by the cameras, encoded and transmitted so that clients can eventually synthesize new images by DIBR techniques.

A population of heterogeneous clients requests camera views from the IMVS system, such that they can freely navigate within the navigation window between views $v_1$ and $v_V$. Navigation is achieved by rendering views in the navigation window possibly with help of DIBR methods. At the decoder side each client can reconstruct a view at any position in the discrete set ${\mathcal{U}}=\{v_1, v_1+\delta, \cdots, V \}$; with $\delta$ as the minimum distance between consecutive views in the navigation window. Due to resource constraints in practical systems, it is not possible to transmit all the camera views to all the clients. Then, clients are clustered according to their bandwidth capabilities and the set of views transmitted to each group of clients are carefully selected, so that their navigation quality is maximized. We propose a \emph{layered multiview representation} with an optimal joint allocation of captured views and coding rates in each layer, which is common for all the clients. The optimization of the layered representation is done centrally by a coordinating engine that we denote as \emph{layered representation creation module} (Fig \ref{fig:Nmodel}).


\subsection{Layered multiview video representation model} \label{subsec:Smodel}

We give now some details on our \emph{layered multiview representation}. The  views $\mathcal{V}$, encoded at rate $\mathcal{R}= \{r_1, r_2, \cdots, r_V\}$, are organized into layered subsets $\mathcal{L} = \{{L}_1, \cdots, {L}_C\}$ to offer a progressive increase of the visual navigation quality with an increasing number of layers. In particular, the finite set of cameras $\mathcal{V}$ is divided in $C$  layers  such that   ${L}_1 \cup {L}_2  \cup \ldots \cup {L}_C \subseteq \mathcal{V} $, with ${L}_i \cap {L}_j=\emptyset, i\neq j$. The number of layers $C$ corresponds to the number of subsets of heterogeneous clients grouped according to their bandwidth capabilities. As a requirement, a client cannot decode a view in ${L}_c$ without receiving the views in ${L}_{c-1}$,  meaning that ${L}_1$ and  ${L}_C$ are the most and the least important subsets, respectively. This means that clients with very low bandwidth capabilities may only receive views in layer ${L}_1$, and need to synthesize the missing viewpoints. On the other hand, clients with higher bandwidth capabilities receive more layers, which leads to a lower rendering distortion as the distance between reference views decreases, hence the view synthesis is of better quality. In addition, we denote by $\mathcal{L}_1^c=\left [L_1, , \cdots, L_c \right ]$ the layers between $L_1$ and $L_c$, and by  $\mathcal{V}_1^c=\left [v_1, \cdots, v_j, \cdots, v_V \right ]$ the ordered subset of views in $\mathcal{V}$ that includes only the views in $\mathcal{L}_1^c$. Here, we assume that view synthesis with DIBR methods is done by using a right and left reference views. Therefore, the leftmost and rightmost views of the navigation window, $v_1$ and $v_V$, need to be transmitted in layer $L_1$. 

Formally, the quality of the interactive navigation when the views from the $c$ most important layers are received and decoded can be defined as:

\begin{align}\label{eq:Dc}
D_c (\mathcal{L}_1^c, \mathcal{R}(\mathcal{V}_1^c))  =  \sum_{\substack{ u\in {\mathcal{U}}, \\  v_l(u), v_r(u) \in  \mathcal{L}_1^c, \\ r_v = \mathcal{R}(v) \forall v \in \{v_l(u), v_r(u)\}   }}     q_u \ d_u (v_l(u), v_r(u))
\end{align}

where ${v}_l(u)$ and ${v}_r(u)$ are the closest right and left reference views to view $u$ among the views in  $\mathcal{V}_1^c$, and $d_u$ is the distortion of view $u$, when it is synthesized using  ${v}_l(u)$ and ${v}_r(u)$ as reference views, encoded at rates $r_v$ in $\mathcal{R}$, for $v$ in $\{v_l(u), v_r(u)\}$. Finally, $q_u$ is the view popularity factor describing the probability that a client selects view  $u \in \mathcal{U}$  for navigation. We assume that $q_u$, depends on the popularity of the views, due to the scene content, but it is independent of  the view previously requested by the client.  
Note that, $D_c \geq D_{c+1}$, since each camera views subset or layer provides a refinement  of the navigation quality experienced by the client.


\subsection{Problem Formulation} \label{sec:pf} 

We now formulate the optimization problem for the allocation of coded views in layers and their rate allocation in order to maximize the expected navigation quality for all IMVS clients. More specifically, the problem is to find the optimal subset of captured views in $\mathcal{V}$ that should be allocated to each of the $C$ layers  $\mathcal{L}^* = \{{L}_1^*, \cdots, {L}_C^*\}$ and coding rate of each selected view, $\mathcal{R^*}= \{r_1^*, \cdots, r_V^*\}$, such that the expected distortion of the navigation is minimized for all the clients, while the bandwidth constraint per layer, $\mathcal{B}=\{B_1, \cdots, B_C\}$, is satisfied. This bandwidth constraint is associated to the bandwidth capabilities of each clients cluster. The optimization of the number of layers and of the rate constraints of the layers due to clients' bandwidth capabilities is out of the scope of this paper. Formally, the optimization problem can be written as:

\begin{equation}\label{eq:LOpt}
 \amin_{\substack{\mathcal{L}, \mathcal{R} \\ |\mathcal{L}|=C, |\mathcal{R}|=V }}   \sum_{c=1}^{C}p(c) \, D_c (\mathcal{L}_1^c, \mathcal{R}(\mathcal{V}_1^c)) 
\end{equation}

such that,

\begin{align}
& & \sum_{v \in \mathcal{V}_1^c} r_v \leq B_c,  & & \forall c \in \{1, \cdots, C\}\nonumber
\end{align}

where $p(c)$ stands for the proportion of clients that are able to receive up to layer ${L}_c$, namely clients with rate capability larger than $B_c$ but lower than $B_{c+1}$. The rate of each encoded view $v$ in $\mathcal{V}_1^c$ is denoted by $r_v$, and the distortion $D_c$ is given in Eq. \ref{eq:Dc}. We finally assume that the depth maps are all encoded at the same high quality, as accurate depth information is important for view synthesis. In practice, the coding rate of depth maps is much smaller than the rate of the texture information, even when compressed at high quality. In the above problem formulation, the rate of encoded views can be formally written as $r_v= r_v^t + r_v^d$, with $r_v^t$ as the rate of the texture information and $r_v^d$ as the rate of the depth information of view $v$. For the sake of clarity, and without loss of generality, we assume in the following that $r_v=r_v^t$, due to the low rate contribution of the compressed depth maps compared with the texture information.

\subsection{NP-Hardness Proof} \label{sec:np}  

We now prove that the optimization problem in (\ref{eq:LOpt}) is NP-hard, by reducing it to a well-known NP-complete problem, the \emph{Knapsack} problem. The Knapsack problem is a combinatorial problem that can be characterized as follows:

\indent \emph{Settings} -- Non-negative weights $w_1, w_2, \cdots, w_V$, profits $c_1, c_2, \cdots, c_V$, and capacity $W$.\\
\indent  \emph{Problem} -- Given a set of items, each with a weight and a profit, find a subset of these items such that the corresponding profit is as large as possible and the total weight is less than $W$.

We now consider a simplified instance of our problem in (\ref{eq:LOpt}) and consider only one layer and a unique rate value for each captured view. Intuitively, if the problem is NP-hard for this simplified case it will also be NP-hard for the full optimization problem. We reduce this simplified problem from the Knapsack problem. First, we map each weight $w_v$ to a view rate $r_v$. Then, when a view $v_j$ is considered as a reference view for the corresponding layer, the profit is quantified by the distortion reduction that it brings, denoted here as $\theta (v_j)$, where $\theta (v_j)= D_c (\mathcal{L}_1^c, \mathcal{R}(\mathcal{V}_1^c)) - D_c (\overline{\mathcal{L}}_1^c, \mathcal{R}(\overline{\mathcal{V}}_1^c))$, for $\overline{\mathcal{V}}_1^c= [\mathcal{V}_1^c \; v_j]$. However, the profit $\theta (v_j)$ of each view is not independent from the content of the layers, as it is the case for each object in the Knapsack problem. The profit depends on the views that have been already selected as reference views in the layer, meaning $\mathcal{V}_1^c$. This increases the complexity of the view selection and rate allocation problem compared to the classic Knapsack problem. Therefore, if the problem is  NP-hard when profits $\theta (v_j)$ are independent of the layer content, it will be NP-hard for our simplified problem. Then, assuming an independent profit for each view, our simplified problem can be rewritten as:

\indent \emph{Settings} -- Rates of the possible reference views $r_1, r_2, \cdots, r_V$, independent profits $\theta (v_1), \theta (v_2), \cdots, \theta (v_V)$, and bandwidth capacity $B_c$.\\
\indent  \emph{Problem} -- Given a set of views, each with a rate and a profit, find the subset of views  such that the distortion reduction is as large as possible and the total rate is less than $B_c$.

This reduced problem is equivalent to the Knapsack problem. Hence, this proves that our original optimization problem is at least as hard as the Knapsack problem. Therefore, our problem in (\ref{eq:LOpt}) is NP-hard.


\section{Proposed Optimization Algorithms}\label{sec:GLA}

To tackle the problem in (\ref{eq:LOpt}), we propose first an algorithm that solves the optimization optimally. Second, we present a reduced complexity algorithm that finds a locally optimal solution working on a layer by layer basis, with an average quality performance close to the optimal algorithm.

\subsection{Optimal Algorithm}\label{sec:GA}

To obtain an optimal solution to the problem in (\ref{eq:LOpt}), we propose a dynamic programming (DP) algorithm that solves problems by breaking them down in subproblems and combining their solutions. The subproblems are solved only once, and their solutions are stored in a DP table to be used in the multiple instances of the same subproblem \cite{Cormen2001}. To develop a DP algorithm from the problem defined in  (\ref{eq:LOpt}), we first need to identify the structure of the problem and how it can be decompose. We start with the following observations:

\begin{enumerate}[leftmargin=0pt,itemindent=1cm]
\item \emph{Decomposition in the view domain} -- We first observe that the rendering quality $D_c$ in Eq. (\ref{eq:Dc}) can be computed by parts. In particular, we can write:

 \begin{equation} \label{eq:Dc_Delta}
D_c (\mathcal{L}_1^c, \mathcal{R}(\mathcal{V}_1^c)) = \sum_{j=1}^{|\mathcal{V}_1^c|-1} \Delta_c(\mathcal{V}_1^c(j), \mathcal{V}_1^c(j+1), \mathcal{R}(\mathcal{V}_1^c(j)), \mathcal{R}(\mathcal{V}_1^c(j+1)))
\end{equation}

 
 where, 
 
 \begin{equation} \label{eq:Delta}
\Delta_c(v_x, v_y, r_x, r_y) =  \sum_{\substack{ u=v_x, \\  v_l(u), v_r(u) \in  \mathcal{V}_1^c, \\ r_i = \mathcal{R}(v_i) \\  \forall v_i \in \{v_l(u), v_r(u)\}}}^ {v_y} q_u \ d_u (v_l(u), v_r(u))
\end{equation} 
 
 is the distortion of rendering views between views $v_x$ and $v_y$, compressed at rates $r_x$ and $r_y$ respectively, using camera views $v_l(u)$ and $v_r(u)$ as the closest left and right reference views of view $u$ in $\mathcal{V}_1^c$. In (\ref{eq:Dc_Delta}), $|\mathcal{V}_1^c|$ stands for the size of the set $\mathcal{V}_1^c$, meaning the number of views in $\mathcal{L}_1^c$.  Note that for the sake of clarity, we drop the parameters $\mathcal{L}_1^c$ and $\mathcal{R}(\mathcal{V}_1^c)$ in the distortion $\Delta_c$, it is however clear that the distortion is only computed for a given layer representation and coding rates. 
 
 Using the above observation, the subset of captured views $\left \{ v_l, v_i, v_r\right \}$, used as reference views, with $v_l < v_i < v_r$, and encoded at rates $\left \{ r_l, r_i, r_r\right \}$, respectively, we can express the distortion  $\Delta_c ( v_l, v_r, r_l, r_r)$ as: 

\begin{equation} \label{eq:Obs1}
 \Delta_c ( v_l, v_r, r_l, r_r) = \Delta_c ( v_l, v_i, r_l, r_i) + \Delta_c ( v_i, v_r, r_i, r_r)
\end{equation}

This decomposition is possible as the distortion in a set of views only depends on the closest right and left reference views for each synthetic view in our system. 

\item \emph{Decomposition in the layer domain} -- Given a multiview layered representation of $C$ layers, the expected distortion between reference views $v_l$ and $v_r$, for the clients groups subscribed to data in layers $c$ to $C$ denoted here as $\phi_c^C\left (  v_l, v_r, r_l, r_r\right ) $  can be expressed as:

\begin{align} \label{eq:Obs2}
 \phi _c^C \left (  v_l, v_r, r_l, r_r  \right ) & = \sum_{i=c}^C p(i) \, \Delta_i ( v_l, v_r, r_l, r_r) \nonumber \\
&=  p(c)  \, \Delta_c ( v_l, v_r, r_l, r_r) +\phi _{c+1}^C\left (  v_l, v_r,r_l,r_r \right )
\end{align}

 As clients receiving higher layers need also to receive all the previous layers for optimal quality improvement, the reference views in layer $c$ become available for any layer $i>c$. This means that the distortion difference for clients in layers $c$ and $c+1$ simply depends on the improvement provided by views in ${L}_{c+1}$. In other words, the expected distortion can be computed iteratively. Finally, note that $\phi_1^C\left (  v_1, v_V, r_1 ,r_V\right ) $ is actually the objective function of our optimization problem in Eq. (\ref{eq:LOpt}), where $\Delta_c ( v_1, v_V, r_1 ,r_V)= D_c (\mathcal{L}_1^c, \mathcal{R}(\mathcal{V}_1^c))$.

\end{enumerate}


Now, let $\Phi _c^C\left ( v_l , v_r, r_l, r_r, \mathcal{B}_c^C \right )$ be the minimum expected distortion $\phi_c^C\left (  v_l,   v_r, r_l, r_r\right ) $, when the rate budget for each layer and for the subset of camera views between $v_l$ and $v_r$ is $ \mathcal{B}_c^C= \left[ B_c, B_{c+1}, \cdots, B_C \right] = \left[ B_c \; \; \mathcal{B}_{c+1}^C\right]$. Based on the above observations, (\ref{eq:Obs1}) and (\ref{eq:Obs2}), this minimum distortion can be defined in a recursive way as follows:


\begin{align}\label{eq:GS}
\Phi_c^C\left ( v_l, v_r, r_l, r_r, B_c^C \right ) = \underset{0 \leq \beta^{C}_{c+1} \leq \mathcal{B}^C_{c+1}} {\underset{0 \leq r_i \leq B_c} { \underset{v_l < v_i < v_r } {\mathrm{min}}}} ~ \; p(c) \Delta_c (v_l, v_i,r_l ,r_i ) + \Phi_{c+1}^C\left ( v_l,  v_i,r_l ,r_i, \beta^{C}_{c+1} \right )  +\Phi _c^C\left ( v_i, v_r, r_i, r_r,  \mathcal{B}_c^C -  \begin{bmatrix}
r_i\\ 
\beta^C_{c+1}
\end{bmatrix} \right ) 
\end{align}

\begin{figure}[h]
\centering
\includegraphics[width=2.6in]{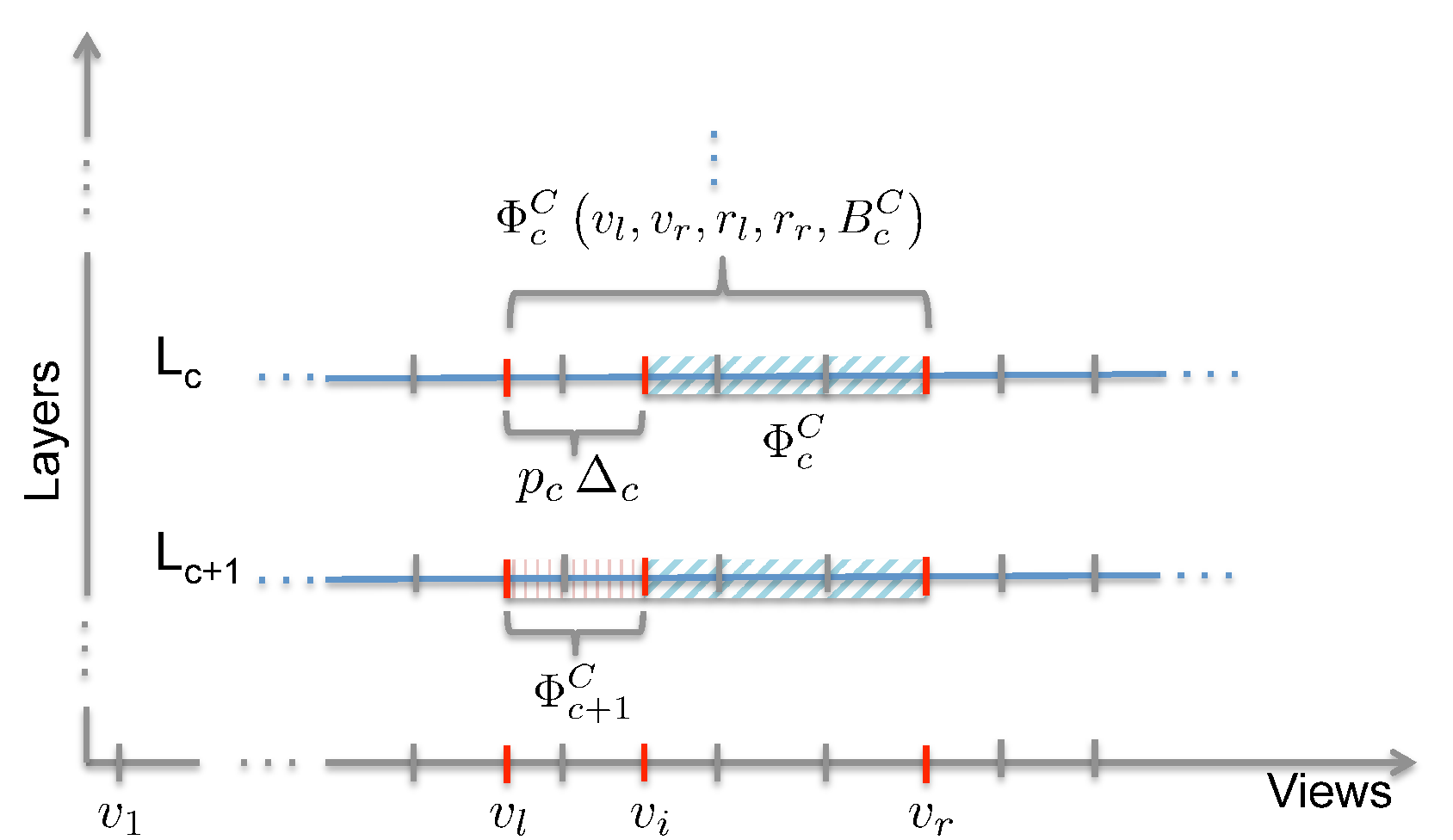}
\caption{Illustration of the optimal algorithm with the three terms from (\ref{eq:GS}) for views in layer $L_c$ to layer $L_{c+1}$ that compose the recursive evaluation of $\Phi _c^C\left ( v_l, v_r, r_l, r_r, B_c^C\right)$.} \label{fig:Optimal}
\end{figure}

The first term in (\ref{eq:GS}) corresponds to the layer distortion $\Delta_c$ between views $v_l$ and $v_i$, as defined in (\ref{eq:Delta}). The second term defines the minimum distortion between views $v_l$ and $v_i$ from layer $c+1$ to layer $C$, when the rate constraint assigned to each layer is $ \beta_{c+1}^C= \left[b_{c+1}, \cdots, b_C \right]= \left[b_{c+1} \; \; \beta_{c+2}^C\right]$, for $\beta_{c+1}^C\leq \mathcal{B}^C_{c+1}$. Finally, the third term is associated to the minimum expected distortion for clients receiving the views from layer $c$ to $C$, between views $v_i$ and $v_r$ when the rate constraint is $\mathcal{B}_c^C -  \begin{bmatrix}
r_i\\ 
\beta^C_{c+1}
\end{bmatrix}$. The recursion relation in (\ref{eq:GS}) directly carries the structure of our DP algorithm that optimally solves the problem in Eq.(\ref{eq:LOpt}). The three terms in (\ref{eq:GS}) are illustrated in Fig. \ref{fig:Optimal} for views from layer $L_c$ to layer $L_{c+1}$.

To solve the problem in (\ref{eq:GS}) using DP algorithm, we follow a bottom-up approach, where the ``smaller'' subproblems are solved first and their solution are used to solve ``larger'' subproblems. In particular the following steps are followed:

\begin{enumerate}
\item Create a  DP table of size $V^2 (B_{max})^{C+2}$, where $B_{max}$ is the maximum bandwidth constraint in $\mathcal{B}_1^C$, and therefore the maximum possible rate value used to encode the camera views. 
\item Start from the last layer $L_C$, with $v_l=v_{V-1}$ and $v_r=v_V$. Solve for every possible encoding rate of $v_{V-1}$ and $v_V$ and rate budget $B_C$. Store each solution in the $(v_{V-1}, v_{V}, \mathcal{R}(v_{V-1}), \mathcal{R}(v_{V}), B_C)$ entry.
\item By progressively moving $v_l$ towards $v_1$, in the same layer, and towards the first layer, following a bottom-up approach, every time $\Phi_c^C$ is called with the argument $(v_l, v_r, r_l, r_r, B_c^C)$, the DP algorithm retrieves the value stored in the corresponding entry of the DP table.
\item Stop when the DP table is completely filled. The optimal solution of the problem is done by comparing the optimal solution in ($v_1$, $v_V$, $r_1$, $r_V$,$B_1$) entries.

\end{enumerate}

The complexity of the DP algorithm that implements (\ref{eq:GS}) can be deduced from the size of the DP table that contains the solutions to the subproblems $\Phi_c^C$, that in this case is $\mathcal{O}(V^2 (B_{max})^{C+2})$. In addition, in order to compute each entry in the DP table, we use (\ref{eq:GS}) where we need at most $(V B_{max})^{CB_{max}}$ comparisons, corresponding to all possible $v_i, \, r_i$ for all possible rate constraints for the following layers $C B_{max}$. Hence the total complexity of the algorithm is $\mathcal{O}\left ( \left (V^{2} (B_{max})^{C+2}\right )^{C B_{max}}\right )$, which is exponential with the number of layers $C$ and the maximum number of rate values used to encode the camera views $B_{max}$.

\subsection{Greedy Algorithm}\label{sec:LA}

The computational time to solve the optimization problem with the above DP algorithm is exponential and rapidly grows with the number of available layers and coding rate values used to encode each of the captured views. 
 Therefore, we propose a greedy approximate solution where the optimization problem defined in (\ref{eq:LOpt}) is solved successively for each layer, starting from the first layer. When solving the optimization problem for each layer successively, the optimal reference views are selected from the full set of captured views when optimizing the first layer, while for the following layers, the solution is restricted to the views that have not been selected as reference views in the previous layers. However, the intuition behind this greedy algorithm is that, in our system, the lowest layers are necessary to most of the clients, for which our greedy algorithm tends to be close to optimal. Therefore, it is expected that our greedy algorithm leads to an effective solution in terms of overall expected distortion. Formally, the greedy algorithm considers the following optimization problem for each layer $L_c$:

\begin{equation}\label{eq:LOptGreedy}
\underset{ L_c,  \mathcal{R}_c  } {\mathrm{min}} \; p(c) \,  D_c (\mathcal{L}_1^c, \mathcal{R}(\mathcal{V}_1^c)) 
\end{equation}

such that,

\begin{equation}
\sum_{v \in \mathcal{V}_1^c} r_v \leq B_c \nonumber 
\end{equation}

where, $\mathcal{R}_c$ stands for the set of coding rates of the views selected as reference views in layer $L_c$.

To obtain an approximate solution, meaning the optimal solution in each particular layer given the set of available reference views, we propose a dynamic programming (DP) algorithm that is inspired on the algorithm in Section \ref{sec:GA}. Let $\Psi_c\left (v_l ,v_V, r_l, r_V, B_c \right )$ be the minimum expected distortion between reference views $v_l$ and $v_V$ compressed at rate $r_l$ and $r_V$, respectively, for clients subscribed to layer $L_c$. The remaining rate budget of $B_c$ is available for selecting new views in layer $L_c$ between the given reference views $v_l$ and $v_V$. This optimal solution is again a recursive function that finds the optimal  $\{v_i, r_i\}$, with $v_l < v_i < v_V$, minimizing $\Delta_c (v_l, v_i, r_l ,r_i)$ and the optimal solution $\Psi_c$ in the remaining set of views between $v_i$ and $v_V$. This can be formally written as:


\begin{equation}\label{eq:LS}
\Psi_c\left ( v_l,  v_V, r_l, r_V, B_c \right ) = \underset{0 \leq r_i \leq B_c} { \underset{v_l < v_i < v_r } {\mathrm{min}}} \;   \Delta_c ( v_l, v_i, r_l,r_i) + \Psi_c\left ( v_i,v_V, r_i ,r_V, B_c - r_i \right )
\end{equation}

A DP algorithm implements the recursive formulation in (\ref{eq:LS}) to determine the optimal allocation of views in layer $L_c$, given the allocation in the lower layers. The algorithm runs for each layer successively, starting from the first layer ${L}_1$. Similarly to the optimal algorithm in Section \ref{sec:GA}, a bottom-up approach is followed for each layer. In particular, the following steps are followed:

\begin{enumerate}
\item Given layer $L_c$, create a  DP table of size $V (B_c)^3$, where $B_c$ is the rate constraint of the layer and therefore the maximum possible rate value to encode the selected reference views. Note that $v_r=v_V$ for every layer; therefore the size of the table regarding the number of views is only $V$, and not $V^2$ as for the optimal algorithm.
\item Start with $v_l=v_{V-1}$ and solve (\ref{eq:LS}) for every possible encoding rate of $v_{V-1}$ and $v_V$ and rate budget $B_c$. Store the solution in the $(v_{V-1}, v_{V}, R(v_{V-1}), R(v_{V}), B_c)$ entry.
\item By progressively moving $v_l$ towards $v_1$ in the same layer, following a bottom-up approach, every time $\Psi_c$ is called with the argument $(v_l, v_V, r_l, r_V, B_c)$, the DP algorithm retrieves the value stored in the corresponding entry of the DP table.
\item Stop when the DP table is completely filled or when $v_l=v_1$. The optimal solution $L_c^*$, $\mathcal{R}_c^*$ of the problem is found by comparing the optimal solution in ($v_1$, $v_V$, $r_1$, $r_V$, $B_c$) entries.
\item If $\mathcal{V}_1^c \subset \mathcal{V}$, and $L_{c+1}$ is available, move to the first step of the algorithm with $L_c = L_{c+1}$. 

\end{enumerate}

\begin{figure}[h]
\centering
\includegraphics[width=2.6in]{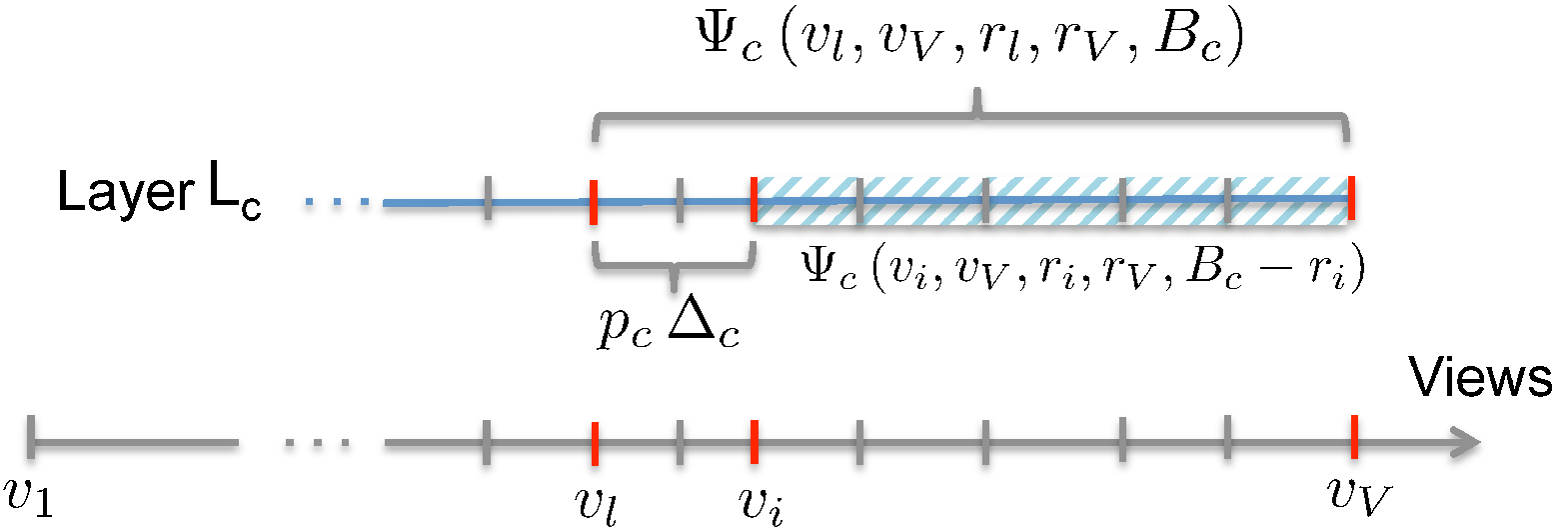}
\caption{Illustration of our greedy algorithm with the two terms from (\ref{eq:LS}) for views $v_l$, $v_i$ and $v_V$ in layer $L_c$.} \label{fig:Approx}
\end{figure}

The different terms in (\ref{eq:LS}) are illustrated in Fig. \ref{fig:Approx} for views $v_l$, $v_i$ and $v_V$ in a particular layer $L_c$ that recursively compute $\Psi _c\left ( v_l,  v_V, r_l, r_V, B_c \right)$. 

Finally, following the same analysis in Section \ref{sec:GA}, $\mathcal{O}\left (C V^{B_{max}+1} (B_{max})^3\right )$ is the complexity of the DP algorithm in (\ref{eq:LS}), as the complexity in solving each of the layers is $\mathcal{O}\left (V^{B_{max}+1} (B_{max})^3\right )$ and the algorithm should run $C$ times, one time for each layer. By solving every layer successively in the greedy algorithm, we are able to remove the exponential dependency with the number of layers, in the complexity of the algorithm; hence to seriously reduce the overall computational complexity of the optimal optimization algorithm.


\section{Performance Assessment}\label{sec:results} 

This section presents the test conditions and performance results obtained in different scenarios when the search of the optimal subset of coded views per layer and rate allocation per view is performed with the algorithms proposed in this paper. We study the optimal allocation in different settings and compare it to the solution of a baseline camera distance-based solution. 


\subsection{ General Test Conditions}\label{sec:results_tc} 

We consider four different data sets for evaluating the performance of our optimization algorithms. We first study the performance on two multiview video datasets, \emph{Ballet} ($1024 \times 768$, 15Hz) \cite{Ballet} and \emph{UndoDancer}  ($1920 \times 1080$, 25Hz) \cite{Dancer}. Though the main target of this work is on video delivery, we also consider multiview image datasets, \emph{Statue} ($2622 \times 1718$) and \emph{Bikes} ($2676 \times 1752$) \cite{Disney}, due to the relatively high quality of their depth maps compared with the ones available in multiview video sequences. Multiview image experiments permits to appreciate the benefits of our solution in allocating resources based on scene content properties. The 3D-HEVC reference software HTM 6.2 \cite{3dhevc} has been used to encode jointly texture and depth maps in each dataset. The views are encoded independently and temporal prediction is used for each view in the video sequences. The depth maps are encoded at high quality (we set a quantizer scale factor of QP=25 for the depth maps), while a set of different rate values $\rho$ is considered for encoding the texture information. For each sequence, the following conditions have been considered:

\begin{itemize}
\item \emph{Statue} -- $V= 7$ captured views and $U=10$ equally spaced rendered views. In this dataset, the cameras are horizontally arranged with a fixed distance between neighboring cameras of 5.33mm. We have chosen the ten available views to have a separation of at least 26.65mm between pair of views, such that $\mathcal{U}= \{50 \;  55 \;  60 \;  65 \;  70 \;  75 \;  80 \;  85 \;  90 \;  95\}$ and $\mathcal{V}= \{50 \;  55 \;  65 \;  70 \;  80 \;  85 \;  95\}$, in terms of view indexes in the dataset.

\item \emph{Bikes} -- $V= 7$ and $U=7$ captured and rendered views, respectively. In this dataset, the cameras are horizontally arranged with a spacing of 5mm. As for \emph{Statue} dataset, to increase the distance between available views, we have chosen the available views by fixing the minimum distance between views to be 25mm. In detail, the seven views correspond to the views $\mathcal{V}= \mathcal{U}=\{10 \;  20 \;  25 \;  30 \;  35 \;  40 \;  50\}$, in terms of dataset indexes. 

\item \emph{Ballet} -- $V= 7$ captured views and $U=8$ rendered views. The views follow a circular arrangement and correspond to $\mathcal{V}=\{0 \;  1 \;  2  \;  4 \;  5 \; 6  \; 7\}$ and $\mathcal{U}=\{0 \;  1 \;  2 \;  3 \;  4 \;  5 \; 6 \; 7 \}$, regarding the view indexes in the dataset.

\item \emph{UndoDancer} -- $V=5$ captured views and $U=9$ equally spaced rendered views.  The cameras for this sequence are horizontally arranged with a fixed distance of 20 cm between neighboring views. They correspond to the captured views $\mathcal{V}=\{1 \;   2 \;  3 \;   5 \; 9\}$ and the nine rendered views $\mathcal{U}=\{1 \;   2 \;  3 \; 4 \;   5 \;  6 \;  7 \;  8 \; 9\}$, in terms of dataset indexes.

\end{itemize}

The distortion of the synthesized view $u$ at the decoder depends on the quality of the reference views used for synthesis, namely $v_l(u)$ and $v_r(u)$, and on their distance to the synthesized view. For the simulations, we use a distortion model proposed in our previous work \cite{Ana2014}, which considers these two factors in estimating the distortion of the synthetic view $d_u$ as:


\begin{equation} \label{eq:Dmodel}
 d_u (v_l, v_r)  =  (1-\alpha)\left ( d_{{v}_{1}^t}(v_l, v_r) + d_{{v}_{1}^d}(v_l, v_r) \right ) +  (1-\gamma)  \alpha \left ( d_{{v}_{2}^t}(v_l, v_r) + d_{\hat{v}_{2}^d}(v_l, v_r) \right ) + \gamma \alpha \mathcal{I}
\end{equation}

where, $d_{{v}_{i}^t}$ and $d_{{v}_{i}^d}$, for $i \in \{1,2\}$, denote the average distortion per pixel for the texture and the depth map of the first and second views that are used as references for view synthesis, where $v_i \in \{v_l(u), v_r(u)\}$. Note that, for the sake of clarity, we have dropped the parameter $u$ from $v_l(u)$ and $v_r(u)$, when it is clear that they refer to the left and right reference view for view $u$ in (\ref{eq:Dmodel}). The parameters $\alpha$ and $\gamma$ are respectively the proportion of disoccluded pixels in the projection of the first reference view and in the projections of both reference views in the DIBR view synthesis. Their values depend only on the scene geometry and they are obtained from the depth maps of the reference views. Finally, the average distortion per pixel in the inpainted areas is denoted by $ \mathcal{I}$, which is assumed to take a constant value that only depends on the scene content. 

In the rest of this section, we carry out simulations for different system settings to evaluate the performance of our greedy and our optimal algorithms presented in Sections \ref{sec:GA} and \ref{sec:LA}. We compare their performance to those of a baseline algorithm, which selects a subset of coded views per layer such that the average distance between reference and synthetic views is minimized in each layer.

%

\subsection{Greedy vs. Optimal algorithm}\label{ssec:results_r1} 

In this section, we compare the performance of both the optimal and greedy algorithms proposed in Sections \ref{sec:GA} and \ref{sec:LA}. Due to the exponential complexity of our optimal algorithm, a small discrete set of available rates $\rho$ to encode the texture information is used and only two layers are considered in the layered multiview representation, which means that the clients are clustered in only two groups depending on their bandwidth capabilities.

We consider two different distributions for the proportion of clients that subscribe to each layer. In particular, we set $p=[0.5  \; 0.5]$, when the first half of the clients can only get ${L}_1$ and the second half get both ${L}_1$ and ${L}_2$, and we set $p=[0.1  \;  0.9]$, when most of the clients have high bandwidth capabilities and only 10\% of them can only get the first layer, ${L}_1$.  We also assume that all the views in $\mathcal{U}$ have the same probability of being requested, which results in a uniform view probability distribution $q$.

The results are presented in Table \ref{tab:LvsG}, where the set of views per layer $\mathcal{L}^*$ and the expected distortion $\overline{D}$, defined as $\sum_{c=1}^{C=2}p(c) D_c$, with $D_c$ in (\ref{eq:Dc}), are shown for each considered data set. The rate constraint per layer $B_c$ and the set of available rates $\rho$ to encode the texture information for each of the considered datasets are given in Table \ref{tab:LvsG}. The views selected by each algorithm in each layer are given in terms of the rate, ${L}_c= \{r_1, \cdots, r_v, \cdots r_V\}$, where $r_v=0$ means that the view is not transmitted in that particular layer and $r_v>0$ means that the view is encoded at rate $r_v$ in the corresponding layer. The indexes of the views correspond to the views arrangement in the set of captured views $\mathcal{V}$.

\begin{table*}[!t]
\renewcommand{\arraystretch}{1.3}
\caption{Comparison of the optimal and greedy algorithms in terms of view selection and rate allocation $\mathcal{L}^*$ and average distortion $\overline{D}$.}
\centering
\begin{tabular}{c|c|c c|c c}
\hline
\bfseries Sequence  & \bfseries Client distribution & \multicolumn{2}{c|}{ \bfseries Optimal}& \multicolumn{2}{c}{ \bfseries Greedy} \\

 \& Settings &  $p$  &  $\mathcal{L}^*$ & $\overline{D} (dB)$ & $\mathcal{L}^*$ & $\overline{D} (dB)$  \\
\hline\hline
\emph{Statue}	& [0.5 0.5] & ${L}_1 =\{2 \; 0  \; 2  \; 0  \; 2  \; 0  \;2\}$ &  29.56 & ${L}_1 =\{2 \; 0  \; 2  \; 0  \; 2  \; 0  \;2\}$ & 29.56  \\
	$B_c=8$Mb		&  & ${L}_2=\{0  \;2 \; 0  \; 2  \; 0  \; 4  \; 0\}$ & & ${L}_2=\{0  \;2 \; 0  \; 2  \; 0  \; 4  \; 0\}$  &  \\
\cline{2-6}
 $\rho= \{ 0 \; 2 \; 4 \} $Mb		& [0.1 0.9] &$L_1 =\{2 \; 0  \; 2  \; 0  \; 2 \; 0  \; 2\}$ & 30.43&  $L_1 =\{2 \; 0  \; 2  \; 0  \; 2  \; 0  \;2\}$ & 30.43 \\
			&  &$L_2=\{0  \;2 \; 0  \; 2  \; 0  \; 4  \; 0\}$ & & $L_2=\{0  \;2 \; 0  \; 2  \; 0  \; 4  \; 0\}$ &  \\
\hline 

\emph{Bikes}	& [0.5 0.5] & $L_1 =\{ 1.5 \;  1.5 \;  0  \; 2  \; 0  \;1.5  \; 1.5 \}$ & 29.00 &  $L_1 =\{ 1.5 \;  1.5 \;  0  \; 2  \; 0  \;1.5  \; 1.5 \}$ & 29.00\\
	$B_c=8$Mb			&  & $L_2=\{0  \; 0  \; 2  \; 0  \; 2  \; 0  \; 0 \}$ & &  $L_2=\{0  \; 0  \; 2  \; 0  \; 2  \; 0  \; 0 \}$&  \\ \cline{2-6}
			$\rho= \{ 0 \; 1.5 \; 2 \} $Mb		& [0.1 0.9] &$L_1 =\{ 2  \;  0 \;  2  \; 0  \; 2  \; 0  \; 2 \}$ & 31.19 &  $L_1 =\{ 1.5 \;  1.5 \;  0  \; 2  \; 0  \;1.5  \; 1.5 \}$ & 29.48  \\
	 &  &$L_2=\{0  \;  2 \; 0 \; 2  \; 0 \;2  \; 0 \}$ & & $L_2=\{0  \; 0  \; 2  \; 0  \; 2  \; 0  \; 0 \}$ &  \\
\hline 

\emph{Ballet}	& [0.5 0.5] & $L_1 =\{ 0.3  \;  0 \;  0  \; 0.3  \; 0  \; 0  \; 0.3 \}$ & 37.33 &  $L_1 =\{ 0.3  \;  0 \;  0  \; 0.3  \; 0  \; 0  \; 0.3 \}$ & 37.33  \\
	$B_c=1$ Mb	&  & $L_2=\{0  \; 0.3  \; 0.3  \; 0  \; 0.3  \; 0  \; 0 \}$ & &$L_2=\{0  \; 0.3  \; 0.3  \; 0  \; 0.3  \; 0  \; 0 \}$ &  \\ \cline{2-6}
		$\rho= \{ 0 \; 0.25 \; 0.3\} $Mb		& [0.1 0.9] &$L_1 =\{ 0.3  \;  0 \;  0  \; 0.3  \; 0  \; 0  \; 0.3 \}$ & 39.27 &  $L_1 =\{ 0.3  \;  0 \;  0  \; 0.3  \; 0  \; 0  \; 0.3 \}$ &39.27 \\
			&  &$L_2=\{0  \; 0.3  \; 0.3  \; 0  \; 0.3  \; 0  \; 0 \}$ & & $L_2=\{0  \; 0.3  \; 0.3  \; 0  \; 0.3  \; 0  \; 0 \}$ &  \\
\hline 

\emph{UndoDancer}	& [0.5 0.5] & $L_1 =\{ 0.5  \; 0  \; 0  \; 1 \; 0.5\}$ & 29.41 & $L_1 =\{ 0.5  \; 0  \; 0  \; 1 \; 0.5\}$  & 29.41\\
$B_c=2$ Mb  &  & $L_2=\{0  \; 1 \; 1 \; 0 \; 0 \}$ & & $L_2=\{0  \; 1 \; 1 \; 0 \; 0 \}$ &  \\\cline{2-6}
	$\rho= \{ 0 \; 0.5 \; 1\} $Mb	& [0.1 0.9] &$L_1 =\{  0.5  \; 0  \; 0  \; 1 \; 0.5\}$ &29.93 & $L_1 =\{ 0.5  \; 0  \; 0  \; 1 \; 0.5\}$  & 29.93 \\
			&  &$L_2=\{ 0  \; 1 \; 1 \; 0 \; 0 \}$ & & $L_2=\{0  \; 1 \; 1 \; 0 \; 0 \}$ &  \\
\hline \hline
\end{tabular}\label{tab:LvsG}
\end{table*}


It can be seen from the results in Table \ref{tab:LvsG} that the same optimal set of views per layer $\mathcal{L}^*$ has been chosen for both the greedy and optimal solutions when a uniform distribution of $p$ for the clients is assumed. The same results have been obtained for values of $p(1)$ higher than 0.5, but they are not presented here due to space restrictions. When $p(2)$ increases, meaning that the second layer $L_2$ is transmitted to a larger group of clients, the greedy algorithm shows its sub-optimality. For instance, when $p(2)=0.9$ the optimal solution is not obtained by the greedy algorithm for the \emph{Bikes} dataset; instead, the same $\mathcal{L}^*$ solution as for $p(1)=p(2)=0.5$ is computed. This sub-optimality is due to the fact that, in our greedy algorithm, the problem is solved successively for each layer, starting from layer $L_1$. This means that the optimal solution $\mathcal{L}^*$ does not depend on the probability distribution $p$ of clients requesting each layer. Therefore, the solution $\mathcal{L}^*$ for each dataset is the same for any distribution $p$; it only affects the expected distortion $\overline{D}$. This successive approach of our greedy algorithm also means that the first layer is prioritized, where the layer $L_1$ always has an optimal set of views independently of the other layers. This explains the good performance of the greedy algorithm  when the first layer has high probability of being transmitted alone, i.e., high value of $p(1)$. Nevertheless, even when the second layer $L_2$ is transmitted to a larger group of clients, $p(2) = 0.9$, the greedy algorithm shows a good performance, presenting an optimal solution for three of the four datasets considered in our experiments. This good performance of the greedy algorithm can be explained by the fact that the first layer is always received by all the clients, independently of the probability distribution $p$ of clients requesting each layer. Therefore, optimizing the allocation of the views in the first layer is never really bad, which further justifies the design of our greedy algorithm. In addition, it has a lower complexity compared with the optimal algorithm, as demonstrated in Section \ref{sec:LA}. Therefore, for the rest of the paper we only consider the greedy algorithm and we compare it with a baseline solution for view selection and rate allocation.

\subsection{Greedy algorithm performance}\label{ssec:results_r2} 

After showing the good performance of our greedy algorithm in the previous section, we now study its performance in different scenarios and compare it with a baseline algorithm, namely \emph{distance-based view selection solution} \cite{Laura2013}. In this algorithm, the views in each layer are selected such that the distance between encoded and synthesized views is minimized. Views are encoded at the same rate in each layer and the rate per view and the number of views are chosen such that the available bandwidth per layer is used to its maximum. Layers are filled in successive order, as for our greedy algorithm.

The algorithms are compared in different settings where the layer rate constraint and view popularity effects are evaluated. A total of four layers are considered in all the simulations presented in this section, representing four groups of clients that are clustered depending on their bandwidth capabilities. Note that, since we do not consider our optimal algorithm in these simulations, we are able to increase the set of available coding rates $\rho$ for each dataset and the number of layers in the multiview layered representation, compared with the experiments in Section \ref{ssec:results_r1}. 


\begin{table}[!t] \footnotesize
\renewcommand{\arraystretch}{1.3}
\caption{ Comparison of the greedy and distance-based algorithm for different layer rate constraints.}
\centering
\begin{tabular}{c| c| c c|c c}
 \hline
 \textbf {Sequence}  & \textbf {Rate constraints} &\multicolumn{2}{c|}{ \textbf {Greedy }}& \multicolumn{2}{c}{ \textbf {Distance-based}} \\
\& Settings & \textbf { \{x, \; y\}}  & $\mathcal{L}^*$ & $ \overline{D} (dB)$ & $\mathcal{L}^*$ & $\overline{D} (dB)$  \\
\hline\hline

\emph{Bikes}	& \{2, 2\}& $L_1 =\{2 \;  0  \; 0  \; 0  \; 0  \; 0  \; 2\}$ & & $L_1 =\{2 \;  0  \; 0  \; 0  \; 0  \; 0  \; 2\} $ & \\ 
	$\rho= \{ 0  \; 1  \; 1.5 \;$	&	& $L_2=\{0  \; 1.5 \; 1.5  \; 0  \; 1.5  \; 1.5 \; 0\}$  & 28.31 & $L_2=\{0  \; 1.5 \; 1.5  \; 0  \; 1.5  \; 1.5 \; 0\}$  &28.31\\ 
	$ 2 \; 2.5  \; 2.7 \}$ Mb			&&   $L_3 =\{0 \; 0  \; 0  \; 2.7  \; 0  \; 0 \; 0\}$ & &   $L_3 =\{0 \; 0  \; 0  \; 2.7  \; 0  \; 0 \; 0\}$ & \\ \cline{2-6}
			& \{0.5, 4\}& $L_1 =\{1.5 \;  0  \; 0  \; 1.5  \; 0  \; 0  \; 1.5\}$ & & $L_1 =\{1.5 \;  0  \; 0  \; 1.5  \; 0  \; 0  \; 1.5\} $ & \\
			&	& $L_2=\{0  \; 2  \; 0 \; 0  \; 1.5  \; 1.5 \; 0\}$  &  28.24  & $L_2=\{0  \; 1.5 \; 0  \; 0  \; 1.5 \; 1.5 \; 0\}$  & 27.92 \\
			&&   $L_3 =\{0 \; 0  \; 2.7  \; 0  \; 0  \; 0 \; 0\}$ & &   $L_3 =\{0 \; 0  \; 0  \; 2.7  \; 0  \; 0 \; 0\}$ & \\
			
\hline

\emph{Ballet}	& \{0.25, 0.25\}& $L_1 =\{0.15 \;  0  \; 0  \; 0.2  \; 0  \; 0  \; 0.15\}$ & & $L_1 =\{0.25 \;  0  \; 0  \; 0  \; 0  \; 0  \; 0.25\} $ & \\
	$\rho= \{ 0  \; 0.15  \; 0.18 \;$		&	& $L_2=\{0  \;0.2 \;0.25  \; 0  \; 0   \; 0.3 \; 0\}$  & 37.33 & $L_2=\{0 \; 0 \; 0.25 \; 0.25    \; 0  \; 0.25 \; 0 \}$  & 36.09  \\
		 $0.20 \; 0.25  \; 0.3 \}$ Mbps	&&   $L_3 =\{0 \; 0  \; 0  \; 0  \; 0.3  \; 0 \; 0\}$ & &   $L_3 =\{0 \; 0.3  \; 0  \; 0  \; 0.3  \; 0 \; 0\}$ & \\ \cline{2-6}
			& \{0.2, 0.1\} & $L_1 =\{0.15 \;  0  \; 0  \; 0 \; 0  \; 0  \; 0.15\}$ & & $L_1 =\{0.15 \;  0  \; 0  \; 0  \; 0  \; 0  \; 0.15\} $ & \\
			&	& $L_2=\{0  \;0.25 \;0  \; 0.25  \; 0   \; 0 \; 0\}$  & 34.91 & $L_2=\{0 \; 0 \; 0.25 \; 0   \; 0.25   \; 0  \; 0 \}$  &34.60  \\
			&&   $L_3 =\{0 \; 0  \; 0.3  \; 0  \; 0  \; 0.3 \; 0\}$ & &   $L_3 =\{0 \; 0.3  \; 0  \; 0  \; 0  \; 0.3 \; 0\}$ & \\
			&&   $L_4 =\{0 \; 0  \; 0  \; 0  \; 0.3  \; 0 \; 0\}$ & &   $L_4 =\{0 \; 0  \; 0  \; 0.3  \; 0  \; 0 \; 0\}$ & \\

\hline

\emph{UndoDancer}	&\{0.5, 0.5\}& $L_1 =\{0.5 \;  0  \; 0   \; 0  \; 0.5\}$ & & $L_1 =\{0.5 \;  0  \; 0   \; 0  \; 0.5\}$ & \\
		$\rho= \{ 0  \; 0.25  \; 0.5 $ 		&	 & $L_2=\{0  \; 0 \; 0.5  \; 1 \; 0 \}$  & 28.64 & $L_2=\{0  \; 0 \; 0.75  \; 0.75  \; 0 \}$  & 28.60  \\
		$\; 0.75 \; 1  \; 1.25 \}$ Mbps		&& $L_3=\{0  \; 1.25  \;0  \;0  \; 0 \}$  & & $L_3=\{0  \; 1.25  \; 0 \; 0  \; 0 \}$  &\\ \cline{2-6}
				&\{0.25, 0.75\}& $L_1 =\{0.5 \;  0  \; 0   \; 0  \; 0.5\}$ & & $L_1 =\{0.5 \;  0  \; 0   \; 0  \; 0.5\}$ & \\
				&	 & $L_2=\{0  \; 0 \; 0  \; 1.25 \; 0 \}$  & 28.26 & $L_2=\{0  \; 0 \; 0  \; 1.25 \; 0 \}$  & 28.26 \\
				&& $L_3=\{0  \; 0.75  \;0.75  \;0  \; 0 \}$  & & $L_3=\{0  \; 0.75  \;0.75  \;0  \; 0 \}$  &\\

\hline \hline
				
\end{tabular}\label{tab:Reffect}
\end{table}

\subsubsection{Layer rate constraint variations}\label{ssec:results_r21} 

In this subsection the greedy algorithm is compared with the distance-based solution in terms of the expected distortion when varying the layer rate constraint. We use an illustrative layer rate distribution that follows a linear relationship: ${B}_c= x \times L_c + y$. By varying the values of $x$ and $y$, we can study the performance of the view selection algorithm in different settings. The corresponding results are presented in Table \ref{tab:Reffect}, where the solution from the greedy algorithm outperforms the distance-based solution in terms of the expected distortion $\overline{D}$ in 4 out of 6 experiments. On the other two cases, the same result is obtained by both algorithm. The performance gain obtained with our greedy algorithm is mainly due to its rate allocation capability compared to the homogeneous rate assignment in the distance-based algorithm. The non-uniform rate allocation of our greedy algorithm permits the fully use of the available rate per layer, allocating more bits to views used as references in the view synthesis process; e.g. for layer $L_2$ with \emph{UndoDancer} sequence when $\{x \; y\}=\{0.5 \; 0.5\}$. In these tests, a distance-based view selection solution shows to be relatively close to the optimal solution, where most of the selected views in each layer are almost equally spaced. This is due to the small change in content among different views, which is due to the small distance between the cameras and/or the low scene complexity in most of the available datasets. Nevertheless, these experiments have shown that a simple distance-based solution with a uniform rate allocation among the selected views in each layer, is not ideal as it cannot take into account the actual content of the scene, contrarily to our algorithm.

\subsubsection{View popularity distribution variations}\label{ssec:results_r22} 

Now we compare our greedy algorithm with the distance-based solution when views have different popularities. The results are shown for an exponential popularity distribution, where the leftmost and rightmost views in the set of captured views $\mathcal{V}$ are the most and the least popular view, respectively. The results are presented in Table \ref{tab:ViewPop}, where the optimal set of views per layer $\mathcal{L}^*$ and the total expected distortion $\overline{D}$ are shown for the greedy and  distance-based solutions. The settings for the different sequences are specified in the Table \ref{tab:ViewPop}. The total expected distortion $\overline{D}$ is calculated assuming a uniform distribution of the proportion clients accessing each layer, meaning $p=[0.25 \; 0.25 \; 0.25 \; 0.25]$, for the four layers. The results show that the solution from the greedy algorithm outperforms the distance-based solution in terms of the total expected distortion. This is due to the fact that the distance-base solution does not consider neither the popularity distribution of the views nor an optimized rate allocation among the views. In particular, in the greedy algorithm the views close to the leftmost view (the most popular views) are selected in the first layers, to ensure that most of the clients receive the most popular views and therefore enjoy a higher expected navigation quality. Similar conclusions can be drawn when considering other view popularities distributions.  

\begin{figure}[!h]
\centering
\subfloat[]{\includegraphics[width=2.6in]{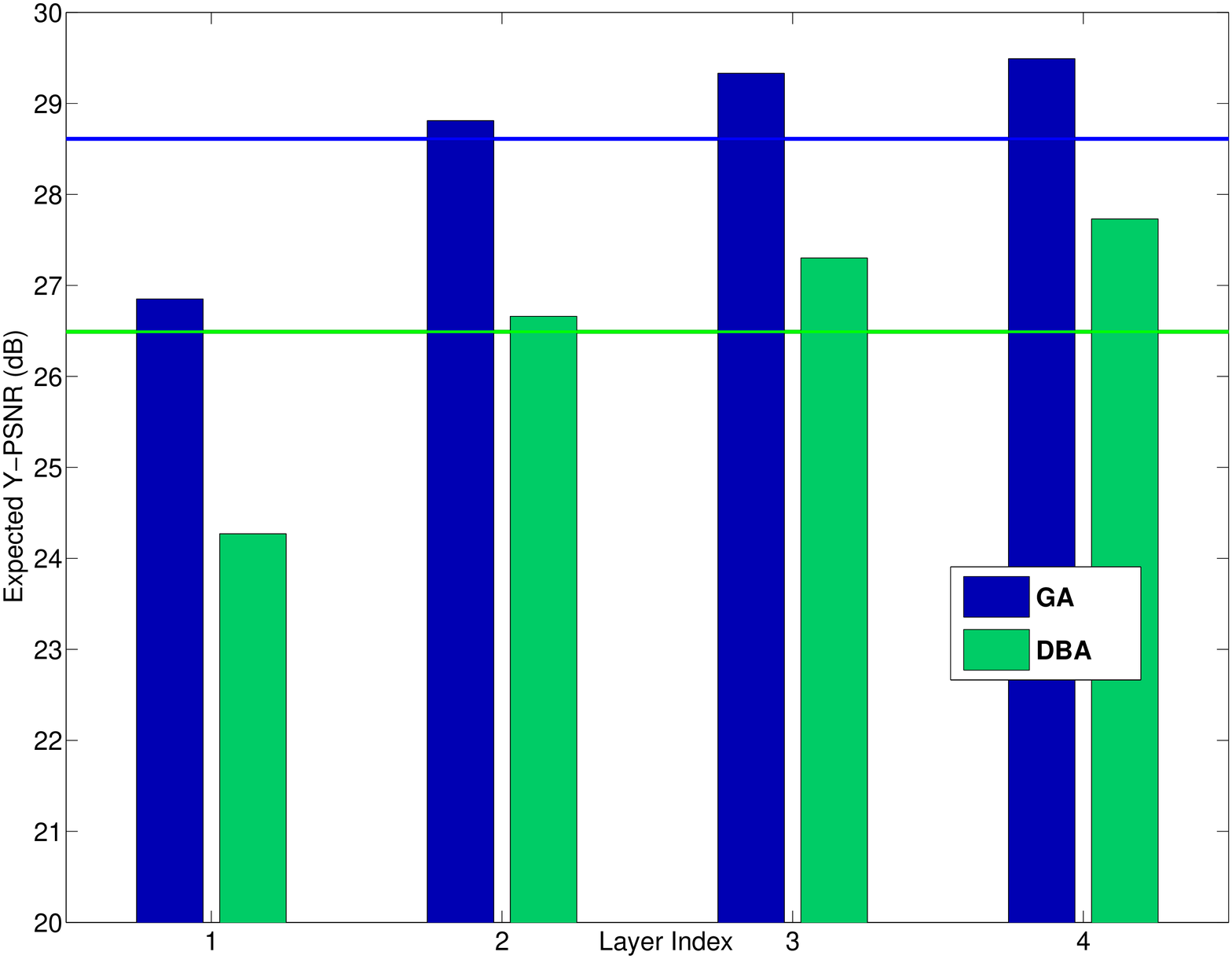}\label{subfig:Exp_effect_Bikes}}\hfil
\subfloat[]{\includegraphics[width=2.6in]{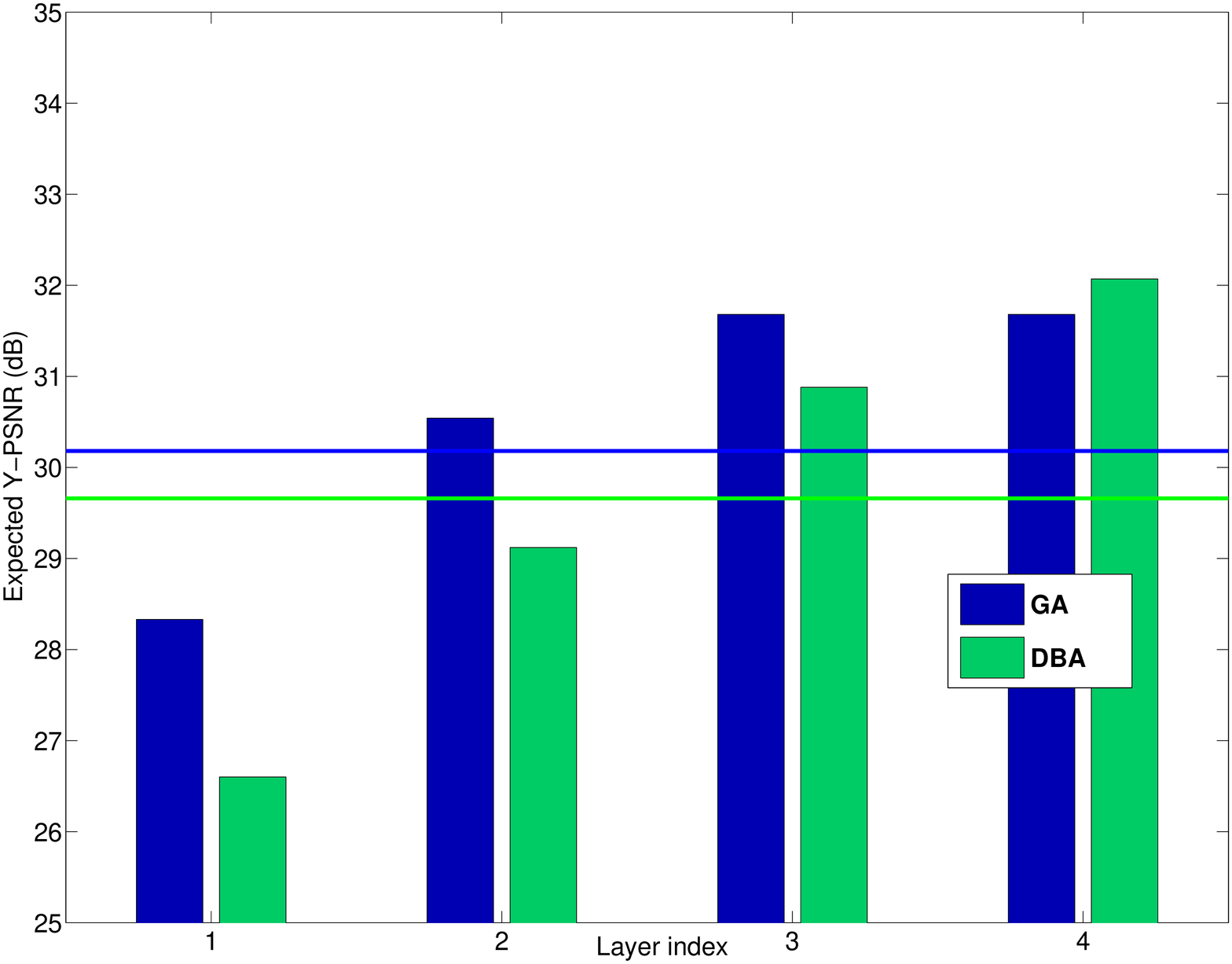}\label{subfig:Exp_effect_UD}}\\
\caption{ Expected Y-PSNR(dB) per client cluster receiving $\mathcal{L}_1^c$, for \emph{Bikes} (a) and \emph{UndoDancer} (b) datasets when comparing our greedy algorithm (GA) and the distance-based algorithm (DBA).}
\label{fig:Exp_effect}
\end{figure}

An alternative presentation of the gain of our greedy algorithm is shown in Fig. \ref{fig:Exp_effect}. A bar plot illustrates the expected quality (Y-PSNR) of our greedy algorithm (GA) and of the distance-based approach (DBA) for the four considered layers in these simulations. We consider the \emph{Bikes} and \emph{UndoDancer} datasets, with the same settings as the ones of the results in Table \ref{tab:ViewPop}. In addition, we have included horizontal lines representing the average quality of each algorithm across the whole client population (the four client clusters), using the same bar color. The distortion is calculated with the views received in the current layer and in all the previous layers, as clients subscribed to a particular layer receive all the views up to that layer. Therefore, for both approaches, the overall quality increases as the layer index increases since clients are able to receive more views. Note that, in general, our greedy algorithm outperforms the distance-based approach, achieving  the highest average quality.  In the case of the \emph{UndoDancer} sequence, we can see however that the group of clients receiving up to layer $L_4$ enjoy a slightly higher quality with the distance-based approach than with the greedy algorithm. This is due to the fact that in layer $L_4$ all the reference views are selected and most of them are encoded at the highest possible rate for the distance-based approach, as it was the only option for the algorithm to fully use the available bandwidth and have a uniform rate allocation among the selected views. However, this view and rate selection of the distance-based solution only favors clients in the last cluster (highest bandwidth capabilities). In fact, the overall performance for the \emph{UndoDancer} sequence is better for our greedy algorithm, as for the first layers the view selection and rate allocation offer a higher quality to the first three group of clients. 

\begin{table*}[!h]
\renewcommand{\arraystretch}{1.3}
\caption{Greedy and distance-based solutions comparison for an exponential view popularity distribution.}
\centering
\begin{tabular}{c| c c|c c}
\hline
\bfseries Sequence & \multicolumn{2}{c|}{ \bfseries Greedy} & \multicolumn{2}{c}{ \bfseries Distance-based} \\

\& Settings & $\mathcal{L}^*$ & $\overline{D} (dB)$ & $\mathcal{L}^*$ & $\overline{D} (dB)$  \\
\hline\hline

\emph{Statue}	& $L_1 =\{4 \;  0  \; 2   \; 0 \; 0  \; 0  \; 0 \; 2\}$& & $L_1 =\{4 \;  0  \;0 \; 0  \; 0  \; 0  \; 0 \; 4\}$& \\
${B}_c= 8 $ Mb	& $L_2=\{0  \; 4 \; 0  \; 0  \; 0  \; 4  \; 0 \; 0\}$  & 34.56 & $L_2=\{0  \; 0 \; 4 \; 0 \; 0  \; 4 \; 0 \; 0\}$  & 34.35 \\
	$\rho= \{ 0  \; 2 \; 4  \;$		&   $L_3 =\{0 \; 0  \; 0   \; 4  \; 4\; 0  \; 0 \; 0\}$ & &   $L_3 =\{0 \; 0  \; 0 \; 4  \; 4 \; 0  \; 0 \; 0\}$ & \\
	$ 5  \; 6 \; 8 \}$ Mb		& $L_4=\{0 \; 0  \; 0 \; 0  \; 0   \; 0\; 4 \; 0\}$ & & $L_4 =\{0 \; 4 \; 0 \; 0  \; 0\; 0  \; 4 \; 0\}$ & \\ \cline{1-5}

\emph{Bikes}	& $L_1 =\{2 \;  0  \; 0  \; 0  \; 0  \; 0  \; 1.5\}$ & & $L_1 =\{1.5 \;  0  \; 0  \; 0  \; 0  \; 0  \; 1.5\} $ & \\
${B}_c= 3.5$Mb				& $L_2=\{0  \; 1.5 \; 2  \; 0  \; 0   \; 0 \; 0\}$  & 28.61 & $L_2=\{0  \; 0 \; 1.5  \; 0  \; 1.5  \; 0 \; 0 \}$  & 26.49 \\
	$\rho= \{ 0  \; 1  \; 1.5 \;$		&   $L_3 =\{0 \; 0  \; 0  \; 2  \; 1.5  \; 0 \; 0\}$ & &   $L_3 =\{0 \; 1.5  \; 0  \; 0 \; 0  \; 1.5 \; 0\}$ & \\
	$ 2 \; 2.5  \; 2.7 \}$ Mb		& $L_4=\{0 \; 0  \; 0 \; 0  \; 0 \; 2.7  \;  0\} $ & & $L_4 =\{0 \; 0  \; 0  \;2.7 \; 0  \; 0 \; 0\}$ & \\ \cline{1-5}

\emph{Ballet}	& $L_1 =\{0.25 \;  0  \; 0  \; 0  \; 0  \; 0  \; 0.25\}$ & & $L_1 =\{0.25 \;  0  \; 0  \; 0  \; 0  \; 0  \; 0.25\} $ & \\
${B}_c= 0.5$Mb	& $L_2=\{0  \;0 \;0.3  \; 0.2  \; 0   \; 0 \; 0\}$  & 37.78 & $L_2=\{0  \; 0 \; 0.25  \; 0  \; 0.25  \; 0 \; 0 \}$  & 37.54 \\
	$\rho= \{ 0  \; 0.15  \; 0.18 \;$		&   $L_3 =\{0 \; 0.3  \; 0  \; 0  \; 0.2  \; 0 \; 0\}$ & &   $L_3 =\{0 \; 0.25  \; 0  \; 0 \; 0  \; 0.25 \; 0\}$ & \\
	$0.20 \; 0.25  \; 0.3 \}$ Mbps		& $L_4=\{0 \; 0  \; 0 \; 0  \; 0 \; 0.3  \;  0\} $ & & $L_4 =\{0 \; 0  \; 0  \;0.3 \; 0  \; 0 \; 0\}$ & \\ \cline{1-5}

\emph{UndoDancer}	& $L_1 =\{0.75 \;  0  \; 0   \; 0  \; 0.5\}$ & & $L_1 =\{0.5 \;  0  \; 0   \; 0  \; 0.5\}$ & \\
${B}_c= 1.25$Mb	 & $L_2=\{0  \; 0.75 \; 0  \; 0.5 \; 0 \}$  & 30.19 & $L_2=\{0  \; 0 \; 0  \; 1.25  \; 0 \}$  & 29.66 \\
$\rho= \{ 0  \; 0.25  \; 0.5 $ 		& $L_3=\{0  \;0 \; 1.25  \;0  \; 0 \}$  & & $L_3=\{0  \; 0 \; 1.25  \; 0  \; 0 \}$  &\\
	$\; 0.75 \; 1 \; 1.25 \}$ Mbps	 & $L_4=\{0  \;0 \; 0  \;0  \; 0 \}$  & & $L_4=\{0  \; 1.25  \; 0  \; 0  \; 0 \}$  &\\

\hline \hline
\end{tabular}\label{tab:ViewPop}
\end{table*}

\section{Conclusion}\label{sec:conclusions} 

We have proposed a novel adaptive transmission solution that jointly selects the optimal subsets of views streams and rate allocation per view for a hierarchical transmission in IMVS applications. We consider a system where the network is characterized by clients with heterogeneous bandwidth capabilities, and we aim to minimize their expected navigation distortion. To do so, clients are clustered according to their bandwidth capabilities and the different camera views are distributed in layers to be transmitted to the different groups of users in a progressive way, such that the clients with higher capabilities receive more layers (more views), hence benefiting of a better navigation quality. We have formulated an optimization problem to jointly determine the optimal arrangement of views in layers along with the coding rate of the views, such that the expected rendering quality is maximized in the navigation window, while the rate of each layer is constrained by network and clients capabilities. To solve this problem, we have proposed an optimal algorithm and a greedy algorithm with a reduced complexity, both based on dynamic-programming. It has been shown through simulations that the proposed algorithms are able to reduce the navigation distortion in a IMVS system. In addition, our greedy algorithm has close-to-optimal performance and outperforms a distance-based algorithm based on an equidistant solution with an uniform rate allocation among the selected views in each layer. Our results show that, considering the client capabilities and their preferences in navigation, and the 3D scene content, as the proposed optimal and greedy algorithms do, is key in the design of an effective adaptive transmission solution for IMVS systems.

\section*{Acknowledgment}

The authors would like to thank Gene Cheung from  the National Institute of Informatics (NII) in Japan for the interesting and fruitful discussions that have substantially enriched this work. This work has been partially supported by \emph{Funda\c{c}\~{a}o para a Ci\^{e}ncia e a Tecnologia}, under the grant SFRH/BD/51443/2011.

\bibliographystyle{elsarticle-num}
\bibliography{VisualComm_2015_elsevier}

\begin{thebibliography}{10}
\expandafter\ifx\csname url\endcsname\relax
  \def\url#1{\texttt{#1}}\fi
\expandafter\ifx\csname urlprefix\endcsname\relax\def\urlprefix{URL }\fi
\expandafter\ifx\csname href\endcsname\relax
  \def\href#1#2{#2} \def\path#1{#1}\fi

\bibitem{Schmeing2011}
M.~Schmeing, X.~Jiang, Depth Image Based Rendering, Springer Berlin Heidelberg,
  2011.
\newblock \href {http://dx.doi.org/10.1007/978-3-642-22407-2\_12}
  {\path{doi:10.1007/978-3-642-22407-2\_12}}.

\bibitem{Ana2014}
A.~De~Abreu, L.~Toni, T.~Maugey, N.~Thomos, P.~Frossard, F.~Pereira, Multiview
  video representations for quality-scalable navigation, in: Proc. of IEEE
  VCIP, Valletta, Malta, 2014.
\newblock \href {http://dx.doi.org/10.1109/PCS.2013.6737710}
  {\path{doi:10.1109/PCS.2013.6737710}}.

\bibitem{Gene2008}
G.~Cheung, A.~Ortega, T.~Sakamoto, Coding structure optimization for
  interactive multiview streaming in virtual world observation, in: Proc. of
  IEEE MMSP, Cairns, Queensland, Australia, 2008.
\newblock \href {http://dx.doi.org/10.1109/MMSP.2008.4665121}
  {\path{doi:10.1109/MMSP.2008.4665121}}.

\bibitem{Gene2011}
G.~Cheung, A.~Ortega, N.-M. Cheung, Interactive streaming of stored multiview
  video using redundant frame structures, IEEE Trans. on Image Processing
  20~(3) (2011) 744--761.
\newblock \href {http://dx.doi.org/10.1109/TIP.2010.2070074}
  {\path{doi:10.1109/TIP.2010.2070074}}.

\bibitem{Gene2012}
X.~Xiu, G.~Cheung, J.~Liang, Delay-cognizant interactive streaming of multiview
  video with free viewpoint synthesis, IEEE Trans. Multimedia 14~(4) (2012)
  1109--1126.
\newblock \href {http://dx.doi.org/10.1109/TMM.2012.2191267}
  {\path{doi:10.1109/TMM.2012.2191267}}.

\bibitem{Liu2010}
Y.~Liu, Q.~Huang, S.~Ma, D.~Zhao, W.~Gao, Rd-optimized interactive streaming of
  multiview video with multiple encodings, J. Vis. Comun. Image Represent.
  21~(5-6) (2010) 523--532.
\newblock \href {http://dx.doi.org/10.1016/j.jvcir.2010.02.004}
  {\path{doi:10.1016/j.jvcir.2010.02.004}}.

\bibitem{Ana2013}
A.~De~Abreu, P.~Frossard, F.~Pereira, Optimized {MVC} prediction structures for
  interactive multiview video streaming, IEEE Signal Processing Letters 20~(6)
  (2013) 603--606.
\newblock \href {http://dx.doi.org/10.1109/LSP.2013.2259815}
  {\path{doi:10.1109/LSP.2013.2259815}}.

\bibitem{Ana2014_journal}
A.~De~Abreu, F.~Pereira, P.~Frossard, Optimizing multiview video plus depth
  prediction structures for interactive multiview video streaming, IEEE Journal
  of Selected Topics in Signal Processing 9~(3) (2015) 487--500.
\newblock \href {http://dx.doi.org/10.1109/JSTSP.2015.2407320}
  {\path{doi:10.1109/JSTSP.2015.2407320}}.

\bibitem{Ana2013_2}
A.~De~Abreu, P.~Frossard, F.~Pereira, Fast {MVC} prediction structure selection
  for interactive multiview video streaming, in: Proc. of PCS, San Jose, CA,
  USA, 2013.
\newblock \href {http://dx.doi.org/10.1109/PCS.2013.6737710}
  {\path{doi:10.1109/PCS.2013.6737710}}.

\bibitem{Takuya2014}
T.~Fujihashi, Z.~Pan, T.~Watanabe, {UMSM}: A traffic reduction method on
  multi-view video streaming for multiple users, IEEE Trans. Multimedia 16~(1)
  (2014) 228--241.
\newblock \href {http://dx.doi.org/10.1109/TMM.2013.2281588}
  {\path{doi:10.1109/TMM.2013.2281588}}.

\bibitem{Gene2011_2}
G.~Cheung, V.~Velisavljevic, A.~Ortega, On dependent bit allocation for
  multiview image coding with depth-image-based rendering, IEEE Trans. on Image
  Processing 20~(11) (2011) 3179--3194.
\newblock \href {http://dx.doi.org/10.1109/TIP.2011.2158230}
  {\path{doi:10.1109/TIP.2011.2158230}}.

\bibitem{Gene2010}
G.~Cheung, V.~Velisavljevic, Efficient bit allocation for multiview image
  coding amp; view synthesis, in: Proc. of IEEE ICIP, 2010, pp. 2613--2616.
\newblock \href {http://dx.doi.org/10.1109/ICIP.2010.5651655}
  {\path{doi:10.1109/ICIP.2010.5651655}}.

\bibitem{SVC}
H.~Schwarz, D.~Marpe, T.~Wiegand, Overview of the scalable video coding
  extension of the {H.264/AVC} standard, IEEE Trans. on Circuits and Systems
  for Video Technology 17~(9) (2007) 1103--1120.
\newblock \href {http://dx.doi.org/10.1109/TCSVT.2007.905532}
  {\path{doi:10.1109/TCSVT.2007.905532}}.

\bibitem{Drose2006}
M.~Drose, C.~Clemens, T.~Sikora, Extending single-view scalable video coding to
  multi-view based on h.264/avc, in: Proc. of IEEE ICIP, 2006, pp. 2977--2980.
\newblock \href {http://dx.doi.org/10.1109/ICIP.2006.312962}
  {\path{doi:10.1109/ICIP.2006.312962}}.

\bibitem{Ozbek2006}
N.~Ozbek, A.~Tekalp, Scalable multi-view video coding for interactive 3{DTV},
  in: Proc. of IEEE ICME, 2006, pp. 213--216.
\newblock \href {http://dx.doi.org/10.1109/ICME.2006.262420}
  {\path{doi:10.1109/ICME.2006.262420}}.

\bibitem{Gene2011_3}
V.~Velisavljevic, V.~Stankovic, J.~Chakareski, G.~Cheung, View and rate
  scalable multiview image coding with depth-image-based rendering, in: Proc.
  of IEEE DSP, 2011, pp. 1--8.
\newblock \href {http://dx.doi.org/10.1109/ICDSP.2011.6005019}
  {\path{doi:10.1109/ICDSP.2011.6005019}}.

\bibitem{Jacob2013}
J.~Chakareski, V.~Velisavljevic, V.~Stankovic, User-action-driven view and rate
  scalable multiview video coding, IEEE Trans. on Image Processing 22~(9)
  (2013) 3473--3484.
\newblock \href {http://dx.doi.org/10.1109/TIP.2013.2269801}
  {\path{doi:10.1109/TIP.2013.2269801}}.

\bibitem{Jacob2015}
J.~Chakareski, V.~Velisavljevic, V.~Stankovic, View-popularity-driven joint
  source and channel coding of view and rate scalable multi-view video, IEEE
  Journal of Selected Topics in Signal Processing 9~(3) (2015) 474--486.
\newblock \href {http://dx.doi.org/10.1109/JSTSP.2015.2402633}
  {\path{doi:10.1109/JSTSP.2015.2402633}}.

\bibitem{Laura2013}
L.~Toni, N.~Thomos, P.~Frossard, Interactive free viewpoint video streaming
  using prioritized network coding, in: Proc. of IEEE MMSP, Pula, Italy, 2013.
\newblock \href {http://dx.doi.org/10.1109/MMSP.2013.6659330}
  {\path{doi:10.1109/MMSP.2013.6659330}}.

\bibitem{Cormen2001}
T.~H. Cormen, C.~Stein, R.~L. Rivest, C.~E. Leiserson, Introduction to
  Algorithms, 2nd Edition, McGraw-Hill Higher Education, 2001.

\bibitem{Ballet}
C.~L. Zitnick, S.~B. Kang, M.~Uyttendaele, S.~Winder, R.~Szeliski, High-quality
  video view interpolation using a layered representation, ACM Trans. on
  Graphics 23~(3) (2004) 600--608.
\newblock \href {http://dx.doi.org/10.1145/1015706.1015766}
  {\path{doi:10.1145/1015706.1015766}}.

\bibitem{Dancer}
D.~Rusanovskyy, P.~Aflaki, M.~M. Hannuksela, Undo dancer {3DV} sequence for
  purposes of {3DV} standardization, in: ISO/IEC JTC1/SC29/WG11 MPEG2010/
  M20028, Geneva, Switzerland, 2011.

\bibitem{Disney}
C.~Kim, H.~Zimmer, Y.~Pritch, A.~Sorkine-Hornung, M.~Gross, Scene
  reconstruction from high spatio-angular resolution light fields, ACM
  Transactions on Graphics 32~(4) (2013) 73:1--73:12.
\newblock \href {http://dx.doi.org/10.1145/2461912.2461926}
  {\path{doi:10.1145/2461912.2461926}}.

\bibitem{3dhevc}
\href{https://hevc.hhi.fraunhofer.de/svn/svn\_3DVCSoftware/tags/HTM-6.2/}{{HTM}
  6.2 software}.
\newline\urlprefix\url{https://hevc.hhi.fraunhofer.de/svn/svn\_3DVCSoftware/tags/HTM-6.2/}

\end{thebibliography}

\end{document}